\documentclass[aps,twocolumn,nofootinbib]{revtex4}
\usepackage{graphicx}
\usepackage{amsmath,amssymb}
\usepackage[colorlinks,citecolor=blue,linkcolor=blue,urlcolor=blue]{hyperref}
\def\be{\begin{equation}}
\def\ee{\end{equation}}
\begin{document}

\title{Planck stars}

\author{Carlo Rovelli}
\affiliation{Aix Marseille Universit\'e, CNRS, CPT, UMR 7332, 13288 Marseille, France.\\
Universit\'e de Toulon, CNRS, CPT, UMR 7332, 83957 La Garde, France.}
  \author{Francesca Vidotto}
 \affiliation{Radboud University Nijmegen, \ Institute for Mathematics, Astrophysics and Particle Physics,
Mailbox 79, P.O. Box 9010, 6500 GL Nijmegen, The Netherlands}

\date{\small\today}

\begin{abstract}
\noindent 
A star that collapses gravitationally can reach a further stage of its life, where quantum-gravitational  pressure counteracts weight.  The duration of this stage is very short in the star proper time, yielding a bounce, but extremely long seen from the outside, because of the huge gravitational time dilation. Since the onset of quantum-gravitational effects is governed by energy density ---not by size--- the star can be much larger than planckian in this phase. The object emerging at the end of the Hawking evaporation of a black hole can then be larger than planckian by a factor $(m/m_{\scriptscriptstyle P})^n$, where $m$ is the mass fallen into the hole, $m_{\scriptscriptstyle P}$ is the Planck mass, and $n$ is positive. We consider arguments for $n=1/3$ and for $n=1$. There is no causality violation or faster-than-light propagation. The existence of these objects alleviates the black-hole information paradox.  More interestingly, these objects could have astrophysical and cosmological interest: they produce a detectable signal, of quantum gravitational origin, around the $10^{-14} cm$ wavelength. 

\noindent \end{abstract}
\maketitle

Measuring effects of the quantum nature of gravity is notoriously difficult \cite{Liberati2011,Hossenfelder2010}, because of the smallness of the Planck scale.  Here we suggest that cosmic rays in the $GeV$ range might contain a trace of a quantum gravitational phenomenon. The large gap between this energy and the Planck energy could be bridged by a large multiplicative factor appearing because of the long (cosmological) lifetime of radiating primordial black holes. This could lead to measurements of ``quantum gravity in the sky'' \cite{Barrau2012}. 

This possibility is suggested by the existence of an apparent paradox currently widely discussed in the theoretical literature. Briefly: what is the fate of the information fallen into the hole, after it evaporates via Hawking radiation? Theoretical arguments appear to indicate that the information being carried out by the radiation implies unpalatable phenomena like ``firewalls'' \cite{Almheiri:2012rt,Braunstein2013}. But if information remains trapped inside, the final stage of the black hole at the end of the evaporation would have to store an amount of information hardly compatible with its expected planckian size \cite{Giddings1992,Page1993,Hossenfelder2010a}.  A possible way out from these unsavory alternatives, suggested by Giddings  \cite{Giddings1992a}, is that the size of the black hole at the final stage of the evaporation is much larger than planckian (see also \cite{Rama2012,Mathur2005}).  

Here we observe that this scenario does not require superluminal transfer of information and can follow from the fact that quantum gravitation phenomena become relevant when the matter energy density reaches the Planck scale, and this may happen at length scales much larger than planckian. 
Quantum gravity may liberate the information stored in the black hole when this is still large compared to the Planck length, implying the existence of a new phase in the life of gravitationally collapsed object \cite{Dai2010}, which could be short in proper time, but, due to gravitational time dilation, very long for an external observer. This, together with the hypothesis of primordial black holes, opens the possibility of measuring a consequence of quantum gravity in cosmic rays. On similar ideas, see also \cite{Bambi2013,Bambi2013a}.

The key insight about the onset of quantum gravitational effects comes first from quantum cosmology. According to loop cosmology \cite{Ashtekar2006}, the Friedmann equation that governs the dynamics of the scale factor $a(t)$ of the universe is modified by quantum gravitational effects as follows
\be
         \left(\frac{\dot a}{a}\right)^2
         =\frac{8\pi G}{3}\rho\left(1-\frac{\rho}{\rho_{\scriptscriptstyle P}}\right), 
\ee
(dot is time derivative, $G$ the Newton constant and $\rho$ the energy density of matter). The quantum correction term in the parenthesis on the r.h.s.\,is determined by the ratio of $\rho$ to a Planck scale density \be
\rho_{\scriptscriptstyle P}\sim{m_{\scriptscriptstyle P}}/l_{\scriptscriptstyle P}^3\sim{c^5/(\hbar G^2)}, 
\ee
where $m_{\scriptscriptstyle P}$ and $l_{\scriptscriptstyle P}$ are the Planck mass and the Planck length, $c$ is the speed of light and $\hbar$ the reduced Planck constant.  Nature appears to enter the quantum gravity regime when the energy density of matter reaches the Planck scale, $\rho\sim\rho_c$. 

The point is that this may happen well before relevant lengths $l$ become planckian ($l$$\sim$$l_{\scriptscriptstyle P}$).  For instance, a collapsing spatially-compact universe bounces back into an expanding one. The bounce is due to a quantum-gravitational repulsion which originates from the Heisenberg uncertainty, and is akin to the ``force'' that keeps an electron from falling into the nucleus \cite{Ashtekar:2006es}.  The bounce does not happen when the universe is of planckian size, as was previously expected; it happens when the matter energy density reaches the Planck density \cite{Rovelli2013e}. In a matter dominated universe this gives a volume of the universe at the bounce 
\be
V\sim\frac{m}{m_{\scriptscriptstyle P}}\ l^3_{\scriptscriptstyle P}
\ee
where $m$ is here the total mass of the universe. Therefore the bounce can happen at large values of the size of the universe: using current  cosmological estimates, quantum gravity could become relevant when the volume of the universe is some 75 orders of magnitude larger than the Planck volume.  Quantum gravity effects do not happen only over Planck volumes.  

The covariant version of the theory \cite{Rovelli2011c} yields the same insight. First, the classical limit of the quantum dynamics is obtained in the regime where two conditions are satisfied: the discrete area quantum number $j$, which is a dimensionless half-integer, must satisfy $j\gg 1$, which is the standard condition for classicality in quantum theory, but it must also be smaller than the curvature radius, that is $j\ll  1/l^2_{\scriptscriptstyle P}R$  where $R$ is the local curvature, in order for the small $\hbar$ limit of the dynamics to converge to general relativity  \cite{Han:2013tap,Han2013a}. The conditions give a bound on $R$, for the validity of the classical theory, and therefore, via Einstein's equations, on the energy density. Second, the theory appears to bound acceleration \cite{Rovelli2013d} with similar consequences. 

\begin{figure}
\centerline{\includegraphics[width=5cm]{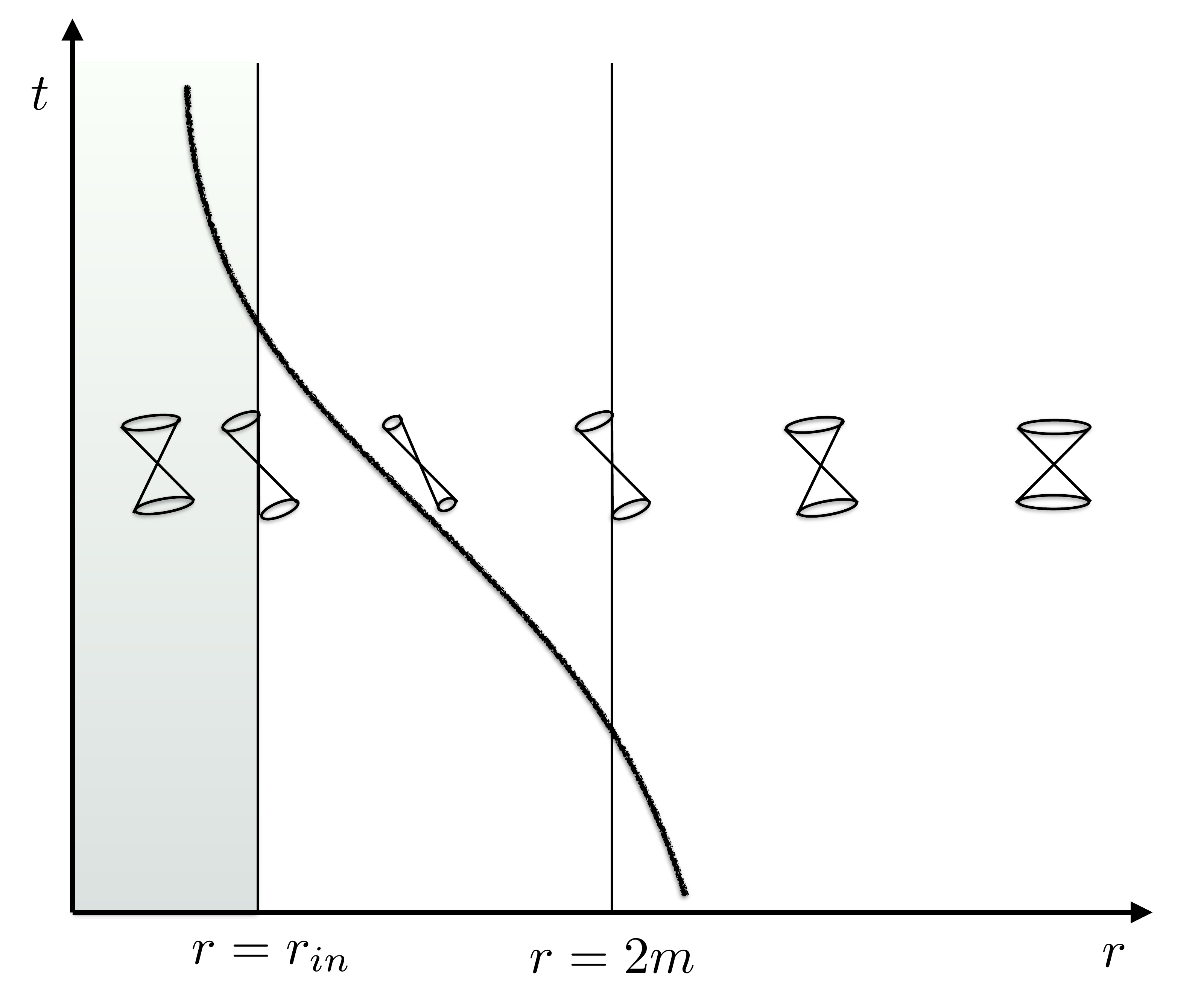}}
\caption{Non evaporating Plank star in Eddington-Finkelstein coordinates. The image illustrates the inward tilting and untitling of the light cones, and a typical infalling timelike geodesic. The quantum gravitational region is shaded.}
\label{ps}
\end{figure}

The analogy between quantum gravitational effects on cosmological and black-hole singularities has been exploited to study if and how quantum gravity could also resolve the $r=0$ singularity at the center of a collapsed star, and there are good indications that it does \cite{Frolov:1981mz,Ashtekar:2005cj,Modesto:2004xx,Modesto2006,Hossenfelder:2009fc,Gambini:2013qf,Goswami2006}.  Thus, consider the possibility that the energy of a collapsing star and any additional energy falling into the hole could condense into a highly compressed core with density of the order of the Planck density \cite{Goswami2006}; see also \cite{Kavic2008b,Kavic2008a}.  If this is the case, the gravitational collapse of a star does not lead to a singularity but to one additional phase in the life of a star: a quantum gravitational phase where the (very large) gravitational attraction is balanced by a (very large) quantum pressure. We call a star in this phase a ``Planck star". Our key observation is that a Planck star can have a size
\be
   r\sim \left(\frac{m}{m_{\scriptscriptstyle P}}\right)^n\  l_{\scriptscriptstyle P}
\ee
where $m$ is now the mass of the star and $n$ is positive.  

For instance, if $n=1/3$ (as in the first naive estimate given below), a stellar-mass black hole would collapse to a Planck star with a size of the order of $10^{-10}$ centimeters.  This is very small compared to the original star --in fact, smaller than the atomic scale-- but it is still more than 30 orders of magnitude larger than the Planck length.  This is the scale on which we are focusing here. The main hypothesis here is that a star so compressed would \emph{not} satisfy the {classical} Einstein equations anymore, even if huge compared to the Planck scale.  Because its energy density is already planckian. 

Can such a Planck star be stable for the lengthy life of a black hole?  The answer is beautiful and surprising (see also \cite{Dai2010}). The life of a Planck star is very long if measured from a distance, because it is determined by the Hawking evaporation time of the black hole in which it is hidden (this is of order $m^3$ in units where $\hbar=G=c=1$ and for a stellar mass black hole is longer than the cosmological time).  But it is extremely short (order $m$, which is the time light takes to cross the radius of the star) if measured on the star itself. The huge difference is due to the extreme gravitational time dilation \cite{Matteo}. Time slows down near high density mass. An observer (capable of resisting the tidal forces) landing on a Planck star will find herself nearly immediately in the distant future, at the time where the black hole ends its evaporation. The \emph{proper} lifetime of a Planck star is short: from its own perspective, the star is essentially a bounce.  A Planck star is a shortcut to the distant future. 

If this is what happens in Nature, the interior of a black hole formed by a star that has collapsed gravitationally might  be modelled by an effective metric that solves the Einstein's equations outside the planckian region, but does not in the inner region, because of quantum effects. The properties of such a metric are easy to visualize: in Eddington-Finkelstein coordinates, the light cones, which bend inwards at the external horizon of the hole, return smoothly pointing upward when entering the Planck region.   Accordingly, there is a second trapping horizon \cite{Hayward1994} inside the Schwarzschild one, at a scale $r_{in}$ related to the size of the Planck star. See Figure \ref{ps}. 

A first naive estimate of the area of the internal trapping surface can be obtained as follows.  On dimensional grounds, the curvature increases as 
\be
R \sim m/r^3
\label{R1}
\ee 
with the radial Schwarzschild coordinate $r$ (defined by $r=\sqrt{A/4\pi}$, where $A$ is the physical area of the constant $r$ sphere); from now on we use natural units where $G=\hbar=c=1$.  Here $m$ is the mass of the collapsed star, which is also the mass of the black hole (after the collapse) measured at infinity, and is connected to the Schwarzschild radius by 
\be
r_{Sch} \sim 2m.
\ee 
From Einstein's equations, curvature is proportional to energy density and our hypothesis is then that when curvature reaches the Planck scale we enter the quantum domain
\be
R \sim 8\pi \rho_{\scriptscriptstyle P}. 
\label{R2}
\ee 
From \eqref{R1} and  \eqref{R2} we have that the boundary of the non-classical region is, neglecting factors of order one, at the radius 
\be
r_{in} \sim \left(\frac{m}{m_{\scriptscriptstyle P}}\right)^{\frac13}\ l_{\scriptscriptstyle P}.
\label{rq}
\ee 
This gives a first naive estimate $n\sim 1/3$. More covariantly, identifying the Planck star with the quantum region, the area of the surface of the Planck star is given by 
\be
A_{q} \sim \left(\frac{m}{m_{\scriptscriptstyle P}}\right)^{\!\frac23} l^2_{\scriptscriptstyle P}.
\label{area}
\ee 
More precise estimates using realistic metrics are of course possible, but do not change much this first estimate \cite{Hossenfelder:2009fc}. 

\begin{figure}
\centerline{\includegraphics[width=5cm]{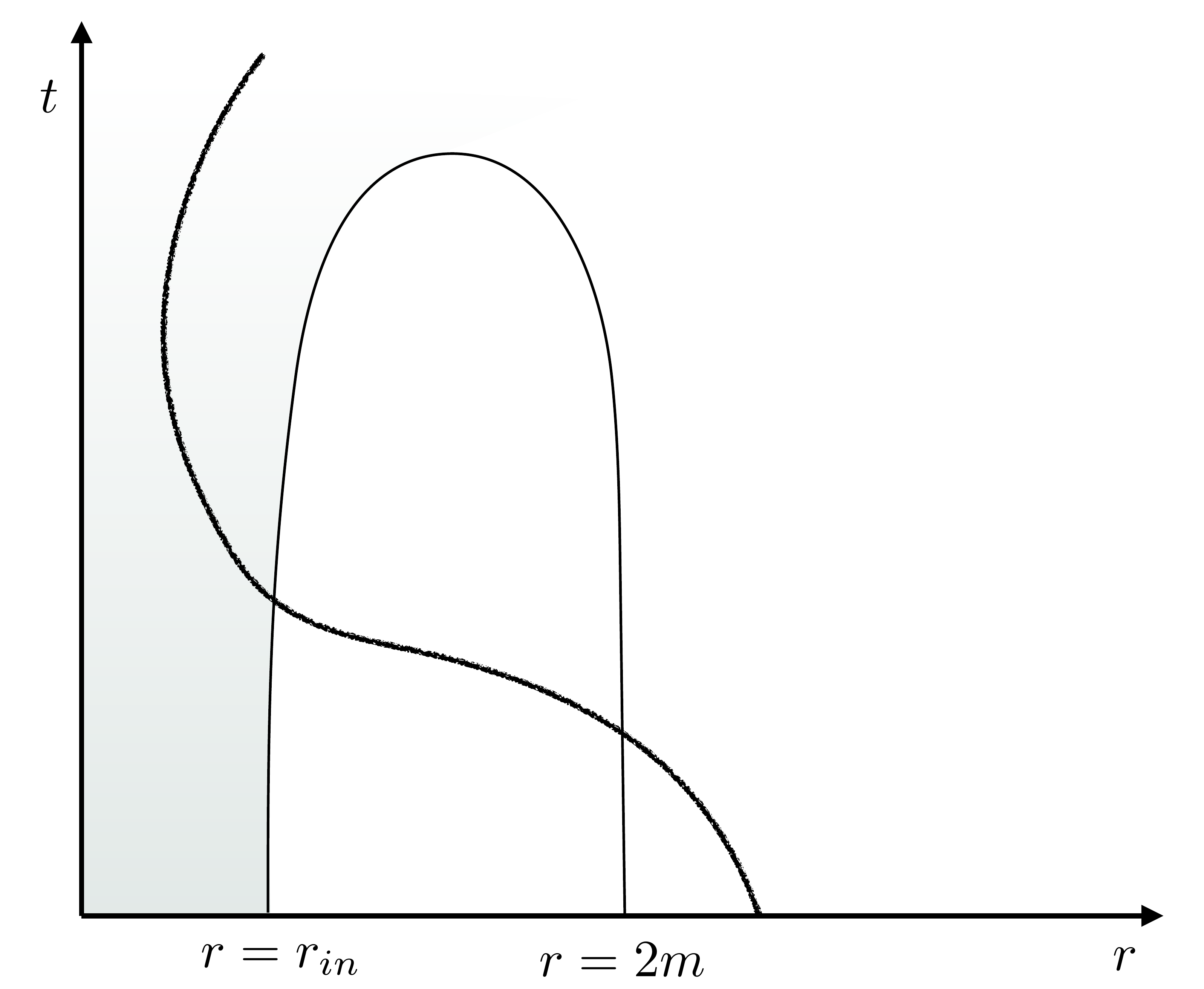}}
\caption{Evaporating Plank star in Eddington-Finkelstein coordinates.}
\label{ps2}
\end{figure}

Let us write a metric that could describe the resulting effective geometry.  In Eddington-Finkelstein coordinates, the metric of a collapsed black hole reads
\be
ds^2=r^2d\omega^2+2dv\,dr-F(r)du^2.
\label{metric}
\ee
where $d\omega^2$ is the metric of a two-sphere and 
\be
F(r)=\left(1-{2m}/{r}\right)
\label{redshift}
\ee
is the standard red shift factor of the Schwarzschild metric. 
The ingoing null geodesics are at constant $u$. The outgoing ones satisfy 
\be
\frac{dr}{dv}=\frac12\left(1-{2m}/{r}\right)
\ee
and therefore are outgoing for $r>2m$ and ingoing for $r<2m$. Once inside the $r=2m$ horizon, a timelike geodesic is bound to hit $r=0$.   We can mimic the quantum gravitational repulsive force by correcting the red shift factor as follows \cite{Hayward2006} 
\be
F(r)=1-\frac{2mr^2}{r^3+2\alpha^2m}
\ee
which gives a regular metric. Expanding in $1/r$ this gives 
\be
F(r)=1-{2m}/{r}+{4 \alpha^2 m^2}/{r^4}.
\ee
The new term represents a strong short-scale repulsive force due to quantum effects. Its effect is to stop the inside bending of the light cones. The ingoing lightlike geodesics turn back vertical at the lowest zero of the term in parenthesis, near \eqref{rq}. (See  \cite{Hayward2006} for a version regular at $r=0$.)  A timelike geodesic that enters the outer horizon will later enter the inner one and then move upward in $t=v-r$ (this $t$ is not the Schwarzschild time coordinate).  See Figure \ref{ps}. 

The inner horizon is at the new lower zero $r_{in}$ of the red shift factor. This happens at $r_{in}\sim \alpha$. For this to be of order $m^n$ (which is our definition of $n$), we must have
\be
\alpha\sim m^{n}.
\ee
For instance, if $n=1/3$ the onset of quantum gravity effects is at the naive $n=1/3$ scale considered above. 

\begin{figure}
\centerline{\includegraphics[width=4.5cm]{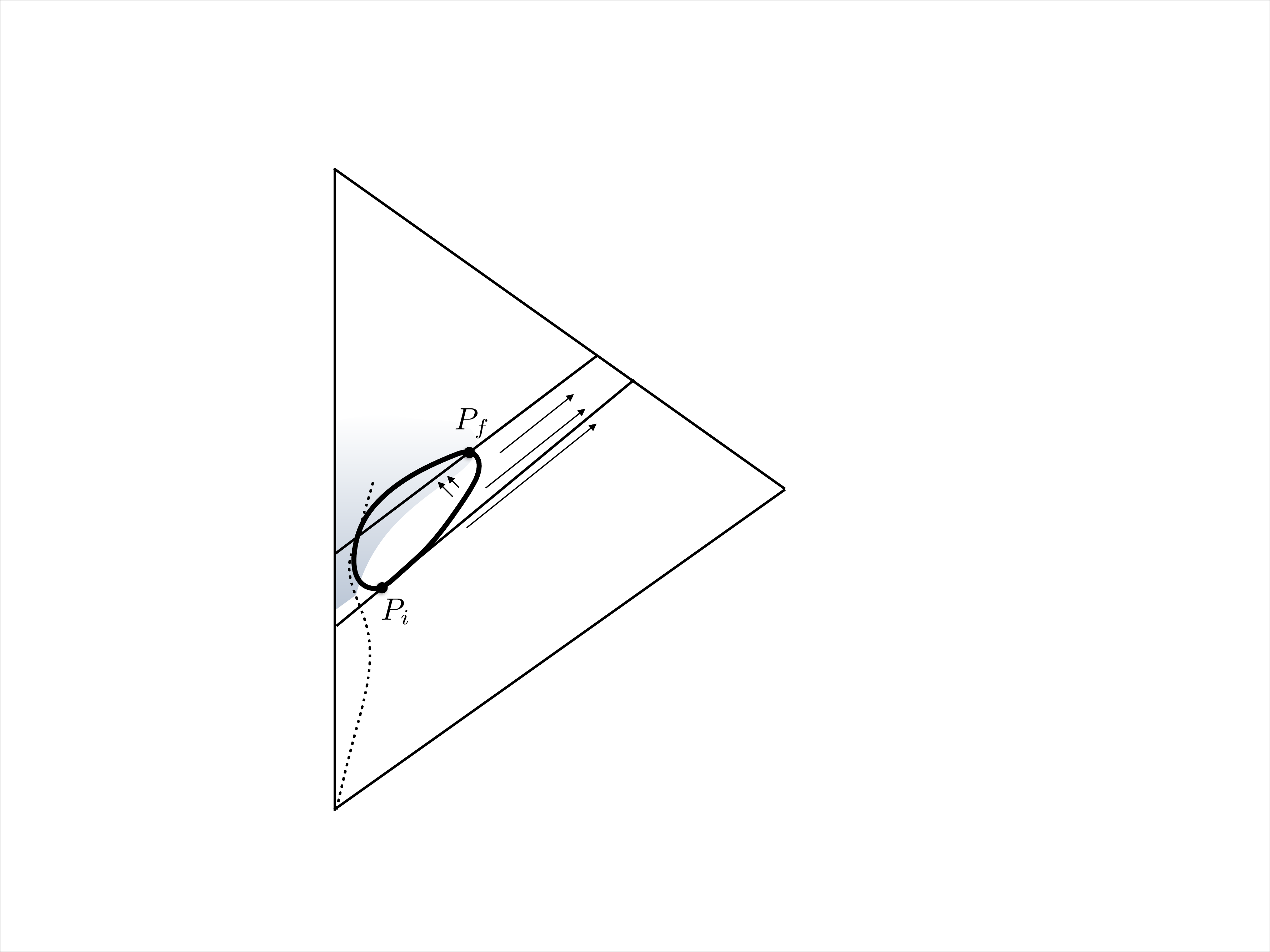}}
\caption{Penrose diagram of the life of a star undergoing gravitational collapse.  The dotted line indicates the external boundary of the star.  The shaded area is the region where quantum gravity modifies the classical Einstein equations.  The dark line represents the two trapping horizons: the external evaporating one, and the internal accreting one. The arrows indicate the Hawking radiation (outgoing and ingoing).  $P_i$ and $P_f$ denote the boundary of the external horizon at the beginning and the end of the Hawking evaporation.  The lowest light-line is where the horizon of the black hole would be if no Hawking evaporation was present.  There is no event horizon.}
\label{ps3}
\end{figure}

Let us now take the Hawking radiation and its back reaction into account. This will give a slow shrinking of the outer horizon, which we can represent in the metric by a $t$ dependence of $m$.  But the inner horizon receives positive energy (because the negative energy partner in a Hawking pair becomes positive energy when in-falling), and therefore will expand. Accordingly, $\alpha$ is going to be time dependent as well, but increasing.

The Hawking radiation slowly shrinks the outer trapping horizon until it reaches the growing internal one. At this point there is no horizon anymore and all information trapped inside can escape. This possibility has been pointed out and studied in \cite{Booth2006} and \cite{Hayward2006}.  This is not going to happen from a Planck size region, but from a macroscopic region, which can contain the residual information that did not escape with the Hawking radiation. The final object of the collapse can be seen as a very short-lived, but \emph{large} remnant, of size much larger than planckian.   This can be illustrated by the metric \eqref{metric} with a red shift factor depending on the time $t=u-r$
\be
F(r,t)=1-{2m(t)}/{r}+{4 \alpha^2(t) m^2}/{r^4},
\ee
where we leave the initial constant $m=m(collapse\ time)$ to emphasise the fact that the Planck star remembers the initial collapsed mass also after the evaporation.   The dependence of $m(t)$ is given by the standard Hawking radiation theory which gives 
\be
\frac{dm(t)}{dt}\sim-\frac{1}{m^2(t)}.
\ee
and the evaporation time \cite{Hawking}
\be
t=\frac{5120\pi G^2 m^3}{\hbar c^4}.
\label{ev}
\ee
In the presence of evaporation, the inner horizon grows, fed by the ingoing partners of the Hawking pairs \cite{Hayward2006} therefore $\alpha(t)$ increases while $m(t)$ decreases. By how much? 

If we assume that the internal quantum region stays essentially the same until the evaporation of the outer horizon reaches it, then the growth of $\alpha(t)$ is small (as in \cite{Hayward2006}) and the final disruption of the black hole is when this has size $(m/m_p)^{\frac13}l_p$. This is already much larger than Planckian, and it is already an interesting result by itself.   This is the first possibility. 

But there is another possibility, which is even more interesting.  We can estimate the final size of the hole  by working backward from the assumption that there are no firewalls. Which is to say, assuming that around the horizon the equivalence principle holds and the quantum state of the fields is not too different from the vacuum.  In terms of order of magnitude this amounts to assuming that the generalsed second law holds at least approximately.  Each outgoing partner of an Hawking pair is maximally entangled with the ingoing one, which falls on the Planck star. Therefore the entropy of the Hawking radiation is purified by the information in the Planck star. The internal entropy cannot exceed the bound given by the area and the Planck star should be sufficiently large to store the corresponding information. It must therefore grow in step with the evaporation of the external horizon shrinking.   If an entropy bound \cite{Bousso2002} holds, when the internal horizon reaches the external horizon, the information in the star must be at least equal to that escaped in the Hawking radiation.  Therefore the area of the surface of the star must, at that point, satisfy
\be
     A_f\sim 4S
\ee
where $S$ is the total entropy of the Hawking radiation emitted up to that point.  But at that point $A$ is also the area of the external horizon, which is four times the residual entropy in the hole.  Thus 
\be
     A_f\sim A_o-A_f\sim 16\pi m^2-A_f
\ee
where $A_o$ was the initial area of the horizon before the Hawking radiation. This gives
\be
A_f= \frac12 A_o\sim 8\pi m^2
\ee
That is
\be
m_f\sim \frac1{\sqrt{2}}\ m
\label{m}
\ee
which is around the Page time \cite{Page1993}. This amounts to $n\sim 1$. This is the second possibility. 

Notice that the evaporation time still remains of the same order of magnitude, because it is proportional to $m^3$; it is only reduced by a factor $\sim .6$.  Therefore for a long period the collapsed star behaves precisely as a conventional black hole.  Nothing changes in conventional black-hole astrophysics. 

The key difference with respect to the conventional scenario which disregards quantum gravity, is that the inner core remembers the original mass.  Approximately one third of the mass is emitted in the Hawking evaporation; at the end of the evaporation, the star is still macroscopic. At this point there is no more horizon, the quantum gravitational pressure can disrupt the star and the information inside the hole can freely escape to infinity.    

The physical picture is compelling: a star collapsing gravitationally can be understood as an object which rapidly shrinks to the size where its energy density is planckian, then bounces back because of the quantum gravitational repulsion due to the quantum properties of spacetime. The bounce takes a short proper time (of the order of $m$, the time light takes to cover the star radius) in the star's own frame.  However, due to the huge gravitational potential, there is a high gravitational redshift that slows the local time with respect to the external world.  An outside observer sees the collapse and the bounce of the star in ``very slow motion", and the entire process takes a long time of order $m^3$.  A black hole is essentially a collapsing and bouncing star that appears frozen because it is seen in slow motion. The information that has fallen into the black hole is just there, frozen by the red shift, waiting to reappear as soon as the bounce is over. 

Notice that there is no violation of causality: in the effective metric considered above, the final disruption of the black hole is in the causal future of the quantum bounce of the star.  The unusual aspects of the solution is a setting-in of quantum effects earlier than expected from simple dimensional considerations or from perturbation theory. 

The existence of such Planck stars can have astrophysical and cosmological implications (see also \cite{Goswami2006,Hossenfelder2012,Kavic2008b,Kavic2008a}). The spectrum of masses of existing black holes is not much understood \cite{Greene2012, Griest2013}.   A possibility that has been repeatedly considered is that a spectrum of primordial black holes with different masses was produced in the early universe; of these, those with mass around $10^{12}\ kg$ have a lifetime of the order of the present age of the universe \cite{Carr1974} 
\be
t_H\sim 14\times 10^{9} years.
\label{t}
\ee
and should therefore be ending their life in our era \cite{Carr2010}, and perhaps be detectable \cite{Cline1992,Cline2011}. If the scenario described in this paper is realistic, their present mass is determined by the age of the universe because the core retains a memory of the initial mass. The corresponding energy could be suddenly liberated today, because of quantum gravitational effects. A black hole of this mass has a size determined by \eqref{ev}, \eqref{m} and \eqref{t}, which give
\be
r^3\sim\frac{G\hbar }{348 \pi c^2}t=\frac{l_{\scriptscriptstyle P}^2 c}{348 \pi}t_H.
\ee
This is 
\be
r=\sqrt[3]{\frac{t_H}{348\pi\ t_{\scriptscriptstyle P}}}\ l_{\scriptscriptstyle P} \sim 10^{-14}\ cm
\ee 
where $t_{\scriptscriptstyle P}=l_{\scriptscriptstyle P}/c$ is the Planck time.  Notice that $r$ is given by the Planck length, which characterizes quantum gravitational phenomena, scaled by a number determined by the ratio of a cosmological scale ($t_H$) to a Planck scale. This ratio is the large multiplicative factor bringing the phenomenon within an observable scale. The sudden dissipation of a star of this size considered in this paper could produce strong signals around this wavelength, namely in the $GeV$ range.  

\centerline{------}

\vspace{.5em}

{\bf Acknowledgments.} Thanks to Stefano Liberati and Eugenio Bianchi for careful comments, to Hal Haggard, Muxin Han, Tommaso De Lorenzo , Goffredo Chirco, Simone Speziale and Alejandro Perez for enlightening discussions. To Don Marolf and Sabine  Hossenfelder for extended discussion on the first draft of this paper. To the participants to the 2014 FQXi conference on Information and in particular Steve Giddings, Andrew Hamilton and Don Page for several exchanges.  CR is particularly grateful to Raphael Busso for a great public discussion.  FV acknowledges support from the Netherlands Organization for Scientific Research (NWO) Rubicon Fellowship Program.  

\bibliographystyle{utcaps}
\bibliography{library}
\end{document}